\documentclass[12pt,twoside]{article}

\usepackage[english]{babel}
\usepackage{fancyhdr}
\usepackage{longtable}
\usepackage{setspace}
\usepackage[mathlines]{lineno}
\usepackage[table]{xcolor}
\usepackage{authblk}
\usepackage{multirow}
\usepackage{amsmath}
\usepackage{graphicx}
\usepackage{epstopdf}
\usepackage{xcolor}
\usepackage{enumitem}
\usepackage{tabularx}
\usepackage{booktabs}
\usepackage[most]{tcolorbox}
\usepackage{xcolor}
\usepackage{caption}
\usepackage{subcaption}
\usepackage{enumitem}
\usepackage{ulem}
\usepackage{wasysym}

\usepackage[numbers,sort&compress]{natbib}

\usepackage{hyperref}
\hypersetup{colorlinks,allcolors=black}
\usepackage[autostyle]{csquotes}
\usepackage{enumerate}
\usepackage{tikz}
\usepackage{amsthm,amssymb}
\usetikzlibrary{shapes.geometric, arrows, positioning}
\usepackage{array}

\newtcolorbox{mybox}[1]{colback=white,colframe=black!40,fonttitle=\bfseries,title=#1}

\newcommand{\cen}[1]{\begin{center} #1 \end{center}}

\begin{document}

\cen{\sf {\Large {\bfseries Medical Physics Dataset Article: 
a database of FLASH Murine In-Vivo Studies} \\  
\vspace*{8mm}
Mathilde Toschini$^{1,2,*}$
Isabella Colizzi$^{1,2, *,\dagger}$
Antony John Lomax$^{1,2}$
Serena Psoroulas$^{1,3}$}\\
$^{*}$Shared first authorship;
$^{1}$Paul Scherrer Institute, Switzerland;
$^{2}$ETH Zurich, Switzerland;
$^{3}$University Hospital Zurich, Switzerland}

\pagenumbering{roman}
\setcounter{page}{1}
\pagestyle{plain}
$\dagger$ Author to whom correspondence should be addressed. Email: isabella.colizzi@psi.ch

\onehalfspacing
\begin{abstract}
\noindent  
{\bf Purpose:} The FLASH effect refers to a lower normal tissue damage for an equivalent tumour response, potentially widening the therapeutic window for radiotherapy. Although this effect has been demonstrated in various experiments using different types of particles and irradiation parameters, the underlying mechanism is not yet clearly understood. Uncertainties surround the conducted experiments, the explored parameter space, and the variability of reported results. To gain a better overview, we have created a dataset that includes in-vivo FLASH experiments. This dataset documents all machine and biological dosimetric parameters, and for determined endpoints, it includes the outcome of the experiment. Our goal with this database is to increase awareness of the results and their variability and provide a useful research and analysis tool for the community.\\
{\bf Acquisition and Validation Methods:} The database contains peer-reviewed papers published until March 2024 on the FLASH in-vivo (murine) experiments. From each paper, previously defined parameters have been manually extracted and/or recalculated to ensure compatibility within the database entries. \\
{\bf Data Format and Usage Notes:} We provide two types of datasets: a user-friendly web-based Notion\cite{noauthor__nodate}  \href{https://exuberant-beak-513.notion.site/354aa5e33bad4771b65168bb613c5768?v=d1046762681f4605af9f4ab756532030&pvs=4}{database} and spreadsheets on a Zenodo repository  \cite{colizzi_isabella_database_2024}. The database contains all the reviewed papers with extracted information in text or numeric form. Users can duplicate the database or view, search, filter, and reorganise online entries. The spreadsheets contain the data for the most analysed endpoints (skin toxicity, survival rate, and crypt cells), allowing a comparative analysis. \\
{\bf Potential Applications:} The study has two main applications. The web-based database will allow for a user-friendly search of information and meta-data of all published FLASH murine data. This will facilitate future research efforts to better understand the FLASH effect. The spreadsheets are a simple and useful tool for the community to conduct statistical analysis and determine the parameters associated with the FLASH effect.
\end{abstract}

\setlength{\baselineskip}{0.7cm}     
\pagenumbering{arabic}
\setcounter{page}{1}
\pagestyle{fancy}

\clearpage
\section{Introduction}
The FLASH effect is a biological effect observed in numerous biological studies when irradiating at ultra-high dose rates (UHDR), a few orders of magnitude higher than those used in routine clinical practice. A precise definition of the effect is still missing and likely involves several interdependent physical and biological parameters \cite{bourhis_clinical_2019}. An analysis of preclinical studies recently suggests that different beam parameters between UHDR and conventional (CONV) dose rates could be responsible for the effect \cite{mcgarrigle_flash_2024}. However, the magnitude of the sparing effect seems to depend on the irradiated tissue \cite{bohlen_normal_2022}.

\noindent
As FLASH researchers, we felt a strong need for an open database encompassing all published studies, explicitly reporting the investigated parameters. In setting up this work, we wanted to understand the parameter space investigated, current gaps, and future directions. Therefore, we paid extensive attention to the completeness of the information reported. Given that this database is the first of its kind, we want to offer an overview of the most frequently reported parameters and propose a standardised approach to reporting experimental data.

\noindent
Additionally, the database can be a useful tool for statistical analysis, helping to understand correlations and dependencies between the different parameters and the FLASH effect - an essential step for successful clinical translation and optimization of the therapeutic window. 

\section{Acquisition and Validation Methods}
\subsection{Review Process}
\begin{figure}[h!]
    \centering
    \begin{tikzpicture}[node distance=2cm]
        \tikzstyle{header} = [rectangle, rounded corners, minimum width= 0.45\textwidth, minimum height=0.55cm, text centered, draw=black,  text width = 0.45\textwidth, font=\footnotesize ]
        \tikzstyle{process1} = [rectangle, minimum width=3cm, minimum height=1cm, text centered, draw=black, fill=orange!20, text width = 0.2\textwidth , font=\footnotesize]
        \tikzstyle{process2} = [rectangle, minimum width=3cm, minimum height=1cm, text centered, draw=black, fill=yellow!20, text width = 0.2\textwidth, , font=\footnotesize]
        \tikzstyle{process3} = [rectangle, minimum width=3cm, minimum height=1cm, text centered, draw=black, fill=red!40, text width = 0.2\textwidth, font=\footnotesize]
        \tikzstyle{arrow} = [thick,->,>=stealth]
        
        \node (header1) [header, fill=orange!40, xshift=-4cm] {\textbf{Identification of studies via database and registers}};
        \node (header2) [header, fill=yellow!40, xshift=4cm] {\textbf{Identification of studies via other methods}};
        
        \node (start1a) [process1, below=0.5cm of header1, xshift=-2cm] {Records identified in PubMed searching for ("FLASH" OR “FLASH-RT” ) AND ("ultra-high" OR "dose-rate") AND ("radiotherapy" OR "radiation" OR "irradiation") AND ("murine" OR "mice" OR “mouse” OR "rat" OR "in-vivo") NOT "review"  \\
        (n = 118)};
        \node (start1b) [process1, below=0.5cm of header1, xshift=2cm] {Records marked as ineligible by filtering (n = 8)};
        \node (start2a) [process2, below=0.5cm of header2, xshift=-2cm] {Records identified from: \\Websites (n = 1);\\  Citation searching (n = 10)};
        \node (process1a) [process1, below=0.5cm of start1a] {Reports screened:\\ (n = 110)};
        \node (process1b) [process1, below=0.5cm of start1a, xshift = 4cm] {Reports excluded:\\ Because of not satisfying criteria (n = 70)};
        \node (process2a) [process2, below=0.5cm of start1a, xshift = 8cm] {Reports screened:\\ (n = 11)};
        \node (process2b) [process2, below=0.5cm of start1a, xshift = 12cm] {Reports excluded:\\ (n = 0)};
        \node (end) [process3, below=0.5cm of process1b, xshift = 2cm] {\textbf{Studies included in the review: \\ (n = 59)}};

        \draw [arrow] (start1a) -- (start1b);
        \draw [arrow] (start1a) -- (process1a);
        \draw [arrow] (process1a) -- (process1b);
        \draw [arrow] (process1a) |- (end);
        \draw [arrow] (start2a) -- (process2a);
        \draw [arrow] (process2a) -- (process2b);
        \draw [arrow] (process2a) |- (end);
    \end{tikzpicture}
    \caption{Simplified version of the PRISMA 2020 flow diagram for systematic reviews \cite{page_prisma_2021}.}
    \label{fig:flow_diagram_PRISMA}
\end{figure}
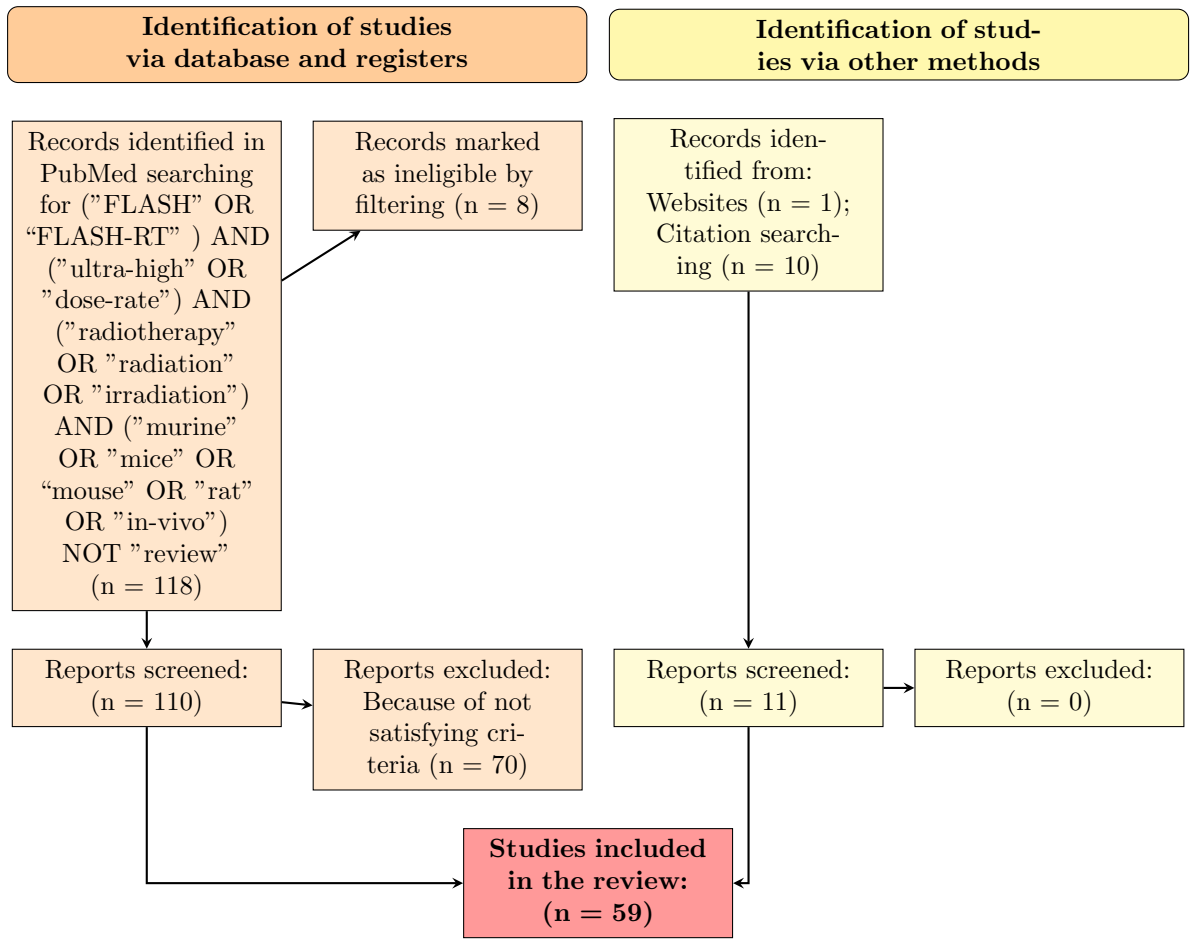

The database collects in-vivo murine publications that compare the UHDR versus CONV dose rate irradiation. Figure \ref{fig:flow_diagram_PRISMA} shows the paper selection process.

\noindent
The inclusion criteria are: 
\begin{itemize}[noitemsep]
    \item Peer-reviewed scientific articles. Conference abstracts, reviews and non-peer-reviewed articles were excluded.
    \item Articles that cover the radiobiological effects of UHDR irradiation on in-vivo models, excluding technical developments regarding UHDR delivery;
    \item Murine (rat or mice) in-vivo studies. Other biological models, such as zebrafish embryos or human patients, were excluded;
    \item Studies comparing the effect of UHDR versus CONV irradiation where the definitions are as outlined in the papers;
    \item Articles about proton, carbon, electron or Xray irradiation;
    \item Articles published from 1966 until March 2024;
    \item Language restricted to English (translations also accepted).
\end{itemize}
\noindent
We searched on PubMed, using the keywords ("FLASH" OR “FLASH-RT” ) AND ("ultra-high" OR "dose-rate") AND ("radiotherapy" OR "radiation" OR "irradiation") AND ("murine" OR "mice" OR “mouse” OR "rat" OR "in-vivo") NOT "review", targeting articles published until March 2024. We identified 118 studies, which we filtered and examined manually, retaining 48 studies satisfying all inclusion criteria. We added 11 papers found through citation searching and websites and ended up with 59 papers.

\subsection{Data extraction and manipulation}
We manually extracted data from the papers, encompassing all parameters provided within them. If specific parameters were not readily available, we sought alternative sources (for instance, previous papers from the same group or vendor information, particularly for machine parameters). When seeking additional information about the biological model or experiment preparation, we contacted the corresponding authors. Whenever possible and necessary, we used standardized definitions to calculate missing parameters; otherwise, we left the field empty. When numerical results were not explicitly provided in the publications or additional materials, we used WebPlotDigitizer \cite{Rohatgi2024} to extract data from original graphs.

\noindent
Most extracted parameters could be included without any particular manipulation. We extracted dosimetric parameters from dose distribution images and calculated them following the ICRU guidelines \cite{noauthor_icru_nodate-1} and \cite{noauthor_icru_nodate}. Due to the different definitions in the literature, we had to recalculate or convert to a common metric to ensure comparability between the data, dose rate for pencil beam scanning (PBS) and endpoint definition for skin toxicity. 

\subsubsection{Computational tool for dose rate calculation for PBS}  \label{DR_code}
Different definitions of dose rates have been proposed for PBS delivery in the literature. To ensure compatibility of the data, we have decided to report the dose rate according to the two definitions that have appeared most frequently: the field average dose rate (delivered dose divided by irradiation time) and the PBS dose rate \cite{folkerts_framework_2020}. If any of the two definitions was missing and was applicable to the analysed paper, we calculated it using information from the paper. In this way, we recalculated the dose rate for all papers consistently. 
The code for PBS dose rate calculation, written in MATLAB, is reported alongside the online database under the page "Dose rate calculation". The calculation works for monoenergetic beams. 

\noindent
In particular, the code allows to calculate:
\begin{itemize}[noitemsep]
    \item the dose distribution
    \item the treatment time and consequent field average dose rate (dose/time)
    \item PBS dose rate of a point of interest within the field
    \item PBS dose rate of all spots and mean PBS dose rate
\end{itemize}
To calculate it, the following parameters are required:
\begin{itemize}[noitemsep]
    \item the beam size in X and Y (assuming a Gaussian distribution)
    \item the spot weight (or amplitude of the Gaussian distribution); specific weight values can be set at the rim or the corners
    \item spot map: number of spots in X and Y and distance between them
    \item timing: spot changing time in X and Y, beam on time per spot (in ms and ms/Gy)
    \item dose level (in percent of the total dose) for contouring the dose distribution and for calculating the field dose rate - for comparability, we used for all studies 95\%
    \item dose threshold (in percent) according to the PBS dose rate metric \cite{folkerts_framework_2020}
    \item if repainting is used, number of repaintings and time between them. The option can also be used for multiple fields. 
\end{itemize}
\noindent
Our calculation was validated against the data reported in the publications. We derived values such as the weight of the single spot by calculating the dose and comparing it to the values reported. Similarly, we could validate the beam-on-time calculation by comparing our field dose rate to the published data.

\subsubsection{Data conversion: skin toxicity}  \label{Skin toxicity}
Skin toxicity is one of the most investigated endpoints. A challenging problem in comparing results for skin toxicity is the difference in the scoring system between the research groups. 
Studies generally distinguish between no or light toxicities, moist desquamation or similar toxicities, and severe toxicities (if any). To allow comparability between the different scales, we have separated the published data into three corresponding groups: in Table \ref{tab:skin_score_scales}, light skin toxicities are represented in yellow, moist desquamation-related toxicities in orange and severe skin toxicities in red. Scoring moist desquamation and the extent of skin it affects is relatively straightforward compared to light skin toxicities, which are more difficult to distinguish. Additionally, severe skin toxicities are generally not allowed in studies due to ethical reasons. Considering these factors, we have converted the experimental data to report the percentage of mice reaching moist desquamation, as this is one of the endpoints that allows a more consistent estimation across different groups.

\section{Data Format and Usage Notes}
This project delivers two essential tools. First, it offers a user-friendly online database housing all relevant papers, categorized by particle and endpoint. The database includes information extracted from various selected parameters. Second, it provides straightforward spreadsheets tailored for statistical analysis for the three endpoints where large datasets could be obtained: crypt cells regeneration, survival rate, and skin toxicity. All paper information has been converted into numerical data in the spreadsheets to enable statistical analysis. Both tools are aligned with the FAIR principle: they are findable through the provided links, accessible without any constraints, interoperable due to the designated system, and reusable, as both can be downloaded or duplicated for further analysis.

\subsection{Notion\cite{noauthor__nodate} Database}
\begin{figure}[h!]
    \centering
    \includegraphics[width=0.9\textwidth]{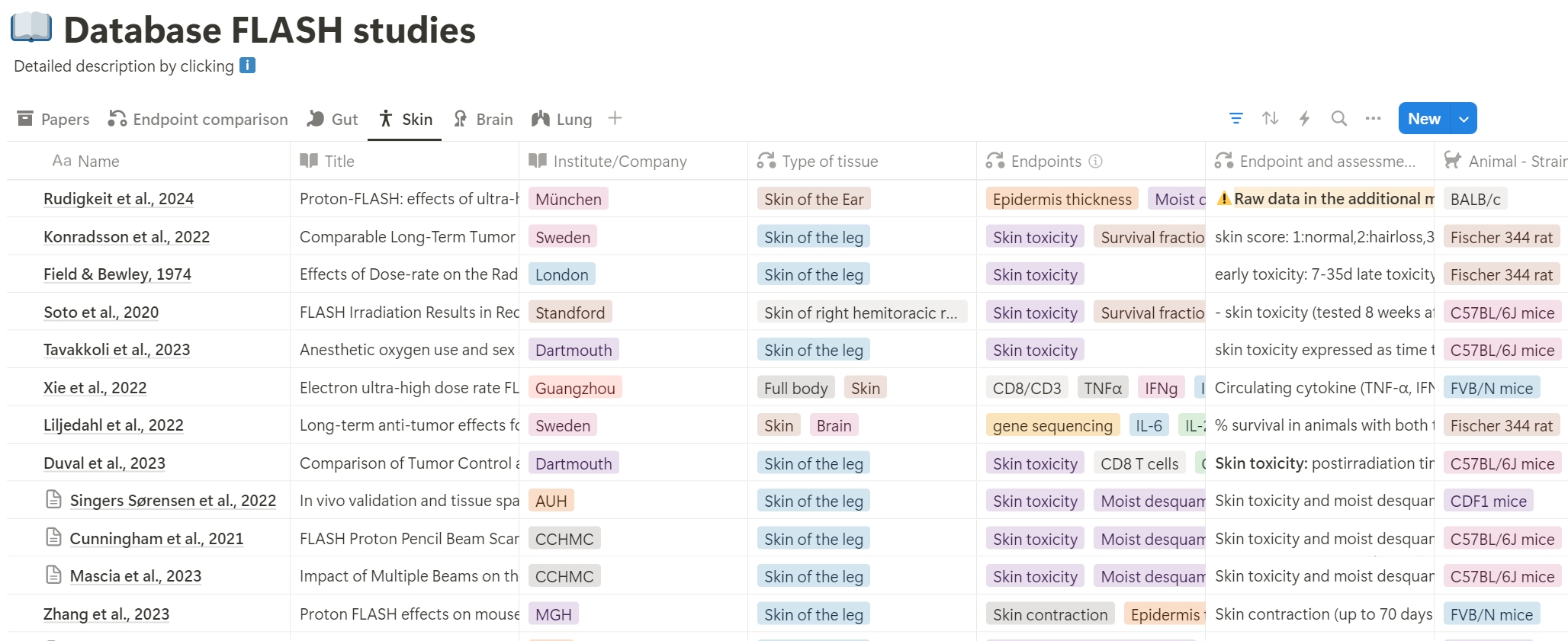}
    \caption{Screenshot of the Notion\cite{noauthor__nodate} database}
   \label{fig: screenshot}
\end{figure}

The database, available at the following \href{https://exuberant-beak-513.notion.site/354aa5e33bad4771b65168bb613c5768?v=d1046762681f4605af9f4ab756532030&pvs=4}{link}, was created using Notion, a productivity and note-taking web application developed by Notion Labs, Inc \cite{noauthor__nodate}. Fig. \ref{fig: screenshot} shows a screenshot. The website accompanying the database consists of an introduction page with the latest updates, a link to the database page, a section with details related to dosimetric and beam parameter calculations, a specific page with commented code for dose rate calculation, a section providing materials for statistical analysis (linked to the Zenodo repository), a simple Python scripts containing data extracted from pictures, and a list of publications where the database has been used. A short video explains how to navigate and use the files for those unfamiliar with Notion\cite{noauthor__nodate}.

\noindent
The website is searchable and allows data to be filtered and grouped. Each article is an entry in the database, with different parameters representing individual attributes. Additional information is provided alongside the parameter name for clarity. Attributes can be multi-select, select, text input, or numerical input. The database has three views, each designed to display specific parameters for better readability and usability. The first view, "Papers", lists all papers based on the particle used, along with their titles and DOIs. In the second view, "Endpoint Comparison", papers are grouped by the endpoint analyzed, providing an overview of the different endpoints and enabling inter-study comparison. The remaining views, "Gut", "Skin", "Brain", and "Lung", display papers filtered for irradiated regions such as the abdominal region, skin (leg, foot, or flank), brain, and lung. These settings are predefined, but users can customize and select specific properties by duplicating the workspace on their own workspace.

\noindent
Table \ref{tab:Parameters} lists all collected parameters with relative descriptions and data types.

\subsection{Spreadsheets for data analysis}
Most of the reported endpoints were evaluated by too few papers, or the way the endpoints were analyzed (time of analysis or procedure) was incompatible. Only three of the data were sufficient for comparison, and we combined them in spreadsheets for statistical analysis. Unlike the Notion\cite{noauthor__nodate} dataset, all the parameters in this dataset are expressed in numerical format. The first page of each spreadsheet provides detailed descriptions of the code used for conversion. The dataset is in Excel XLSX \cite{msexcel} format, simplifying the scripting process, and uploaded to a Zenodo repository \cite{colizzi_isabella_database_2024}.

\subsubsection{Spreadsheet \textit{FLASH\_parameters\_Gut}}
This spreadsheet reports the outcomes of experiments performed irradiating the abdominal region. It collects data from two endpoints: survival rate and number of regenerated crypts. The dataset is divided into:
\begin{itemize}
    \item \textit{gut\_crypt} reports results based on the percentage of regenerating crypt after irradiation.
    \item \textit{gut\_crypt\_effect} consider the FLASH effect as either a continuous variable, ranging between 0 and 1, with 0 indicating no effect, 1 indicating UHDR (ultra-high dose rate) being better than CONV (conventional), and -1 indicating CONV being better than UHDR. The effect is calculated as:
    \begin{equation}
            \text{Effect}_{crypt.} = \dfrac{P_{UHDR} - P_{CONV}}{100}
        \label{eq: effect}
    \end{equation}
    where $P_{UHDR}$ and $P_{CONV}$ represent the percentages of survival crypt cells after UHDR and CONV irradiation, respectively.  If the absolute number of crypt cells was reported, we assumed that the average number per investigated area was 140 \cite{martin_altered_1998}.   
    \item \textit{gut\_survival\_\%} and \textit{gut\_survival\_01} report the percentage of mice surviving on day 11 or day 20 after irradiation or indicate whether each mouse is alive (0) or dead (1).   
    \item \textit{gut\_effect\%} and \textit{gut\_effect01} consider the FLASH effect as either a continuous variable or a discrete variable, with 0 indicating no effect, 1 indicating UHDR being better than CONV, and -1 indicating CONV being better than UHDR. The effect is calculated as in \ref{eq: effect}. This calculation is done for both day 11 and day 20.
\end{itemize}

\subsubsection{Spreadsheet \textit{FLASH\_parameters\_Skin}}
This spreadsheet reports the outcomes of experiments evaluating the skin after irradiation, converted to the common grading scheme as described in \ref{Skin toxicity}. The dataset is divided into:
\begin{itemize}
    \item \textit{Max\_Moist\_desquamation\_percent} reports the percentage of mice that reached moist desquamation after irradiation
    \item \textit{Max\_Moist\_desquamation\_effect\%} and \textit{Max\_Moist\_desquamation\_effect01} consider the FLASH effect as either a continuous variable, ranging between 0 and 1, or a discrete variable, with 0 indicating no effect, 1 indicating UHDR (ultra-high dose rate) being better than CONV, and -1 indicating CONV being better than UHDR. The effect is calculated as:

    \begin{equation}
        \text{Effect}_{M.d.}= \dfrac{P_{CONV} - P_{UHDR}}{100}
    \label{eq: effect md}
    \end{equation}
    $P_{UHDR}$ and $P_{CONV}$ represent the percentages of mice reaching moist desquamation after UHDR and CONV irradiation, respectively. 
\end{itemize}

\section{Discussion}                                                           

For the first time, we created an open web-based database encompassing published studies, providing comprehensive information on dosimetry, biological effects, and machine-specific parameters, among other key data points. 
In addition to the web-based database, we provide the community access to spreadsheets containing numerical and scriptable information for three important endpoints (crypt cell count, survival rate, and moist desquamation). These can be used for statistical analysis to identify parameters responsible for the FLASH effect.  

\noindent
Most parameters could be successfully extracted from publications, delivering a nearly complete database.Some parameters were omitted, such as the dosimetry instrument used and the irradiated volume, since these were not always reported or could not be inferred. It is important to address the variability and incompleteness of the data by examining each specific parameter group:
\begin{itemize}[noitemsep]
    \item Biological parameters:
        The sex and age of the rodents were not always provided, and similarly, the oxygen level during the irradiation process was not well-documented. While we assume the irradiation to be under normal oxygen conditions, it has been found that the oxygen level during anaesthesia can significantly affect the treatment outcome \cite{iturri_oxygen_2023}.
    \item Endpoints:
        The endpoint analysis and time of analysis were reported. However, analysis criteria were often different between papers, as in the case of skin toxicity. Reporting a survival curve implies the existence of criteria to determine the point of termination of the animal, which can vary between labs and depend on ethical guidelines; this information was sometimes missing. When interpreting data, careful attention must be paid to differences in endpoint evaluation. Finally, many endpoints are analysed by only a few or one group.  
    \item Machine parameters:
        Some publications did not provide all the necessary details about the machines used, particularly the settings for CONV irradiation. Additionally, most beam parameters were reported at the source instead of the target position. Information could be found in previously published papers describing the FLASH "commissioning" of the machine used; we therefore assumed that the parameters are the same as those mentioned in the previous paper. This is particularly relevant for pencil beam scanning, where the correct information on the number of spots, spot map, spot size, and spot changing time is necessary for recalculating the dose rate.         
    \item Dosimetric parameters:
        Many publications do not provide dosimetric data, such as dose uniformity and lateral and distal fall-off, so the database could not be completed. Additionally, the definition of prescribed dose is not unique across different irradiation modalities. It was often not specified over which isodose the dose was defined or how homogenous the dose distribution was. This is particularly important for papers with only a point dose measurement.
         
\end{itemize}
\noindent
Many parameters were extracted from figures, introducing uncertainties due to the inherent nature of data extraction. We strongly advocate providing raw data with publications to ensure higher precision in meta-analyses. Additionally, incorporating histograms\cite{harrison_novel_2024} that display dose, dose rate, and linear energy transfer, as well as the design of the energy modulator utilized, would enhance the reevaluation and comparison of preclinical data. As the last limitation of our database, we note that the selection of parameters to report and the definitions used were based on common practice, as no guideline was present and required us to calculate missing parameters when needed. A recent proposal for standardization will hopefully improve the situation in the future.

\noindent
To further advance our understanding of the FLASH effect, it is crucial to increase accessibility to more experimental studies. Keeping a comprehensive database of relevant publications that can be easily maintained is vital for this purpose. Compared to other databases, we found the web-based Notion\cite{noauthor__nodate} database to be user-friendly and intuitive. It has been beneficial for streamlining the database creation process, allowing for robust querying, flexible parameter grouping, and the addition of detailed descriptions, images, and files. Additionally, it supports seamless collaboration among multiple users and enables easy website creation. These features would allow authors of new papers focusing on FLASH in vivo studies to contribute directly to the database, minimizing maintenance efforts.

\noindent
In the last years, there has been significant interest in the FLASH effect. However, we identified the lack of a comprehensive open database containing all published data. We therefore decided to provide the community with a web-based database that will allow easy access to information and metadata on all published data related to FLASH, and an open data repository. Even though authors already shared much information in current publications, we also identified a strong need for consistency in definitions and reporting methods, especially regarding dosimetry and biological endpoints. Establishing a consensus within the community on what parameters should be reported and how is crucial. We hope this work will stimulate discussions and guide the community toward a future agreement.

\section{Conclusion}

We developed an online database with all in-vivo FLASH studies published up to March 2024. The gathered data will streamline the statistical analysis process and collectively advance future research efforts to better understand the FLASH effect. It will also assist in designing future experiments, enhance compatibility between research centres, and increase researchers' awareness of the parameters that need to be reported.

\section{Aknowledgment}
The authors thank Robert Schäfer for providing computational support. We also want to express our gratitude to the FLASH community and all the corresponding authors and contributors who helped us fill the gaps in the database. We are especially thankful to Brita Sørensen, Jan Schuemann and Beth Rothwell for sharing additional data and engaging in stimulating discussions about future experiments and practices. \\
This work was funded by the Swiss National Science Foundation (Grant No. 200882) 

\section{Author contribution}
M.T.: data curation, analysis, conceptualisation, writing (original draft);
I.C.: supervision, data curation, analysis, conceptualisation, writing (original draft);
S.P.: supervision, conceptualisation, funding acquisition, project administration, writing (review and editing);
T.L.: funding acquisition, project administration, writing (review and editing).

\nolinenumbers

\clearpage

\begin{thebibliography}{10}

\bibitem{noauthor__nodate}
© 2024 {Notion} {Labs}, {Inc}.

\bibitem{colizzi_isabella_database_2024}
Isabella Colizzi, Mathilde Toschini, Antony~J. Lomax, and Serena Psoroulas.
\newblock A database of {FLASH} {Murine} {In}-{Vivo} {Studies}, February 2024.

\bibitem{bourhis_clinical_2019}
Jean Bourhis, Pierre Montay-Gruel, Patrik Gonçalves~Jorge, Claude Bailat, Benoît Petit, Jonathan Ollivier, Wendy Jeanneret-Sozzi, Mahmut Ozsahin, François Bochud, Raphaël Moeckli, Jean-François Germond, and Marie-Catherine Vozenin.
\newblock Clinical translation of {FLASH} radiotherapy: {Why} and how?
\newblock {\em Radiotherapy and Oncology}, 139:11--17, October 2019.

\bibitem{mcgarrigle_flash_2024}
Josie~May McGarrigle, Kenneth~Richard Long, and Yolanda Prezado.
\newblock The {FLASH} effect—an evaluation of preclinical studies of ultra-high dose rate radiotherapy.
\newblock {\em Frontiers in Oncology}, 14, April 2024.
\newblock Publisher: Frontiers.

\bibitem{bohlen_normal_2022}
Till~Tobias Böhlen, Jean-François Germond, Jean Bourhis, Marie-Catherine Vozenin, Esat~Mahmut Ozsahin, François Bochud, Claude Bailat, and Raphaël Moeckli.
\newblock Normal {Tissue} {Sparing} by {FLASH} as a {Function} of {Single}-{Fraction} {Dose}: {A} {Quantitative} {Analysis}.
\newblock {\em International Journal of Radiation Oncology, Biology, Physics}, 114(5):1032--1044, December 2022.

\bibitem{Rohatgi2024}
Ankit Rohatgi.
\newblock Webplotdigitizer, 2024.

\bibitem{noauthor_icru_nodate-1}
{ICRU} {Report} 71, {Prescribing}, {Recording}, and {Reporting} {Electron} {Beam} {Therapy} – {ICRU}.

\bibitem{noauthor_icru_nodate}
{ICRU} {Report} 78, {Prescribing}, {Recording}, and {Reporting} {Proton}-{Beam} {Therapy} – {ICRU}.

\bibitem{folkerts_framework_2020}
Michael~M. Folkerts, Eric Abel, Simon Busold, Jessica~Rika Perez, Vidhya Krishnamurthi, and C.~Clifton Ling.
\newblock A framework for defining {FLASH} dose rate for pencil beam scanning.
\newblock {\em Medical Physics}, 47(12):6396--6404, 2020.

\bibitem{msexcel}
{Microsoft Corporation}.
\newblock Microsoft excel.

\bibitem{martin_altered_1998}
K.~Martin, C.~S. Potten, S.~A. Roberts, and T.~B. Kirkwood.
\newblock Altered stem cell regeneration in irradiated intestinal crypts of senescent mice.
\newblock {\em Journal of Cell Science}, 111 ( Pt 16):2297--2303, August 1998.

\bibitem{iturri_oxygen_2023}
Lorea Iturri, Annaïg Bertho, Charlotte Lamirault, Elise Brisebard, Marjorie Juchaux, Cristèle Gilbert, Julie Espenon, Catherine Sébrié, Laurène Jourdain, Ludovic de~Marzi, Frédéric Pouzoulet, Jane Muret, Pierre Verrelle, and Yolanda Prezado.
\newblock Oxygen supplementation in anesthesia can block {FLASH} effect and anti-tumor immunity in conventional proton therapy.
\newblock {\em Communications Medicine}, 3(1):1--13, December 2023.
\newblock Number: 1 Publisher: Nature Publishing Group.

\bibitem{harrison_novel_2024}
Nathan Harrison, Minglei Kang, Ruirui Liu, Serdar Charyyev, Niklas Wahl, Wei Liu, Jun Zhou, Kristin~A. Higgins, Charles~B. Simone, Jeffrey~D. Bradley, William~S. Dynan, and Liyong Lin.
\newblock A {Novel} {Inverse} {Algorithm} {To} {Solve} the {Integrated} {Optimization} of {Dose}, {Dose} {Rate}, and {Linear} {Energy} {Transfer} of {Proton} {FLASH} {Therapy} {With} {Sparse} {Filters}.
\newblock {\em International Journal of Radiation Oncology, Biology, Physics}, 119(3):957--967, July 2024.
\newblock Publisher: Elsevier.

\bibitem{page_prisma_2021}
Matthew~J. Page, Joanne~E. McKenzie, Patrick~M. Bossuyt, Isabelle Boutron, Tammy~C. Hoffmann, Cynthia~D. Mulrow, Larissa Shamseer, Jennifer~M. Tetzlaff, Elie~A. Akl, Sue~E. Brennan, Roger Chou, Julie Glanville, Jeremy~M. Grimshaw, Asbjørn Hróbjartsson, Manoj~M. Lalu, Tianjing Li, Elizabeth~W. Loder, Evan Mayo-Wilson, Steve McDonald, Luke~A. McGuinness, Lesley~A. Stewart, James Thomas, Andrea~C. Tricco, Vivian~A. Welch, Penny Whiting, and David Moher.
\newblock The {PRISMA} 2020 statement: an updated guideline for reporting systematic reviews.
\newblock {\em BMJ}, 372:n71, March 2021.
\newblock Publisher: British Medical Journal Publishing Group Section: Research Methods \&amp; Reporting.

\bibitem{sorensen_pencil_2022}
Brita~Singers Sørensen, Mateusz~Krzysztof Sitarz, Christina Ankjærgaard, Jacob~G. Johansen, Claus~E. Andersen, Eleni Kanouta, Cai Grau, and Per Poulsen.
\newblock Pencil beam scanning proton {FLASH} maintains tumor control while normal tissue damage is reduced in a mouse model.
\newblock {\em Radiotherapy and Oncology}, 175:178--184, October 2022.

\bibitem{mascia_impact_2023}
Anthony Mascia, Shelby McCauley, Joseph Speth, Stefanno~Alarcon Nunez, Gael Boivin, Marta Vilalta, Ricky~A. Sharma, John~P. Perentesis, and Mathieu Sertorio.
\newblock Impact of {Multiple} {Beams} on the {FLASH} {Effect} in {Soft} {Tissue} and {Skin} in {Mice}.
\newblock {\em International Journal of Radiation Oncology*Biology*Physics}, August 2023.

\bibitem{velalopoulou_flash_2021}
Anastasia Velalopoulou, Ilias~V. Karagounis, Gwendolyn~M. Cramer, Michele~M. Kim, Giorgos Skoufos, Denisa Goia, Sarah Hagan, Ioannis~I. Verginadis, Khayrullo Shoniyozov, June Chiango, Michelle Cerullo, Kelley Varner, Lutian Yao, Ling Qin, Artemis~G. Hatzigeorgiou, Andy~J. Minn, Mary Putt, Matthew Lanza, Charles-Antoine Assenmacher, Enrico Radaelli, Jennifer Huck, Eric Diffenderfer, Lei Dong, James Metz, Constantinos Koumenis, Keith~A. Cengel, Amit Maity, and Theresa~M. Busch.
\newblock {FLASH} {Proton} {Radiotherapy} {Spares} {Normal} {Epithelial} and {Mesenchymal} {Tissues} {While} {Preserving} {Sarcoma} {Response}.
\newblock {\em Cancer Research}, 81(18):4808--4821, September 2021.

\bibitem{field_effects_1974}
S.B. Field and D.K. Bewley.
\newblock Effects of {Dose}-rate on the {Radiation} {Response} of {Rat} {Skin}.
\newblock {\em International Journal of Radiation Biology and Related Studies in Physics, Chemistry and Medicine}, 26(3):259--267, January 1974.

\bibitem{konradsson_comparable_2022}
Elise Konradsson, Emma Liljedahl, Emma Gustafsson, Gabriel Adrian, Sarah Beyer, Suhayb~Ehsaan Ilaahi, Kristoffer Petersson, Crister Ceberg, and Henrietta Nittby~Redebrandt.
\newblock Comparable {Long}-{Term} {Tumor} {Control} for {Hypofractionated} {FLASH} {Versus} {Conventional} {Radiation} {Therapy} in an {Immunocompetent} {Rat} {Glioma} {Model}.
\newblock {\em Advances in Radiation Oncology}, 7(6):101011, November 2022.

\bibitem{soto_flash_2020}
Luis~A. Soto, Kerriann~M. Casey, Jinghui Wang, Alexandra Blaney, Rakesh Manjappa, Dylan Breitkreutz, Lawrie Skinner, Suparna Dutt, Ryan~B. Ko, Karl Bush, Amy~S. Yu, Stavros Melemenidis, Samuel Strober, Edgar Englemann, Peter~G. Maxim, Edward~E. Graves, and Billy W.~Loo Jr.
\newblock {FLASH} {Irradiation} {Results} in {Reduced} {Severe} {Skin} {Toxicity} {Compared} to {Conventional}-{Dose}-{Rate} {Irradiation}.
\newblock {\em Radiation Research}, 194(6):618--624, August 2020.
\newblock Publisher: Radiation Research Society.

\bibitem{duval_comparison_2023}
Kayla E.~A. Duval, Ethan Aulwes, Rongxiao Zhang, Mahbubur Rahman, M.~Ramish Ashraf, Austin Sloop, Jacob Sunnerberg, Benjamin~B. Williams, Xu~Cao, Petr Bruza, Alireza Kheirollah, Armin Tavakkoli, Lesley~A. Jarvis, Philip~E. Schaner, Harold~M. Swartz, David~J. Gladstone, Brian~W. Pogue, and P.~Jack Hoopes.
\newblock Comparison of {Tumor} {Control} and {Skin} {Damage} in a {Mouse} {Model} after {Ultra}-{High} {Dose} {Rate} {Irradiation} and {Conventional} {Irradiation}.
\newblock {\em Radiation Research}, 200(3):223--231, August 2023.
\newblock Publisher: Radiation Research Society.

\bibitem{rudigkeit_proton-flash_2024}
Sarah Rudigkeit, Thomas~E. Schmid, Annique~C. Dombrowsky, Jessica Stolz, Stefan Bartzsch, Ce-Belle Chen, Nicole Matejka, Matthias Sammer, Andreas Bergmaier, Günther Dollinger, and Judith Reindl.
\newblock Proton-{FLASH}: effects of ultra-high dose rate irradiation on an in-vivo mouse ear model.
\newblock {\em Scientific Reports}, 14(1):1418, January 2024.
\newblock Publisher: Nature Publishing Group.

\bibitem{tavakkoli_anesthetic_2023}
Armin~D. Tavakkoli, Megan~A. Clark, Alireza Kheirollah, Austin~M. Sloop, Haille~E. Soderholm, Noah~J. Daniel, Arthur~F. Petusseau, Yina~H. Huang, Charles~R. Thomas, Lesley~A. Jarvis, Rongxiao Zhang, Brian~W. Pogue, David~J. Gladstone, and P.~Jack Hoopes.
\newblock Anesthetic oxygen use and sex are critical factors in the {FLASH} sparing effect.
\newblock {\em bioRxiv: The Preprint Server for Biology}, page 2023.11.04.565626, November 2023.

\end{thebibliography}

\listoffigures
\listoftables

\clearpage
\section{Tables}

\small
\begin{longtable}{p{5cm}|p{8cm}|p{2cm}}
    \caption{List of all collected parameters with relative descriptions and data types. The parameters are grouped per category.} \\
    
    \hline
    \rowcolor{lightgray} \multicolumn{3}{c}{\textbf{General Information}} \\
    \textbf{Parameter} & \textbf{Description} & \textbf{Type} \\
    \hline
    \endfirsthead
    \multicolumn{3}{c}{\textit{Continued from previous page}} \\
    \hline
    \textbf{Parameter} & \textbf{Description} & \textbf{Type} \\
    \hline
    \endhead
    \hline
    \multicolumn{3}{c}{\textit{Continued on next page}} \\
    \endfoot
    \endlastfoot

    Title & Title of the paper & Text \\
    Year & Year of publication & Number \\
    DOI & Digital Object Identifier of the paper & Text (link) \\
    In-text and Full Citation & Short and long citation format & Text \\
    Abstract & Full abstract as provided online & Text \\
    Authors & List of authors and coauthors & Text \\
    Institute/Company & Institute/Company to which the first author belongs to & Multi-select \\

    \hline
    \rowcolor{lightgray} \multicolumn{3}{c}{\textbf{Biological Model}} \\
    Sample size & Number of rodents included in the trial. Separation between CONV and FLASH (if given) & Number \\
    Rodents strain & Strain of the mice/rat & Multi-select \\
    Rodents age [w] & Age (in weeks) at the time of irradiation & Number \\
    Rodents sex & Female or Male & Multi-select \\
    Rodents additional information & Information regarding irradiation region, specificity of the mice or additional information regarding the tumour & Text \\
    Anaesthesia & Type of anesthesia used & Multi-select \\
    Oxygen level & Oxygen level at which the mice were irradiated & Multi-select \\
    Depilation & Depilation method, only relevant for skin irradiation & Multi-select \\
    Tumour and tumour type & Presence or Absence of the tumour and if present, type of tumour cells & Multi-select \\

    \hline
    \rowcolor{lightgray} \multicolumn{3}{c}{\textbf{Endpoint}} \\
    Type of Tissue & Irradiated body area & Multi-select \\
    Endpoints & List of endpoints analysed & Multi-select \\
    Endpoint and Assessment time & Additional information regarding the endpoint, type and timepoint of analysis and metric used (if given) & Text \\
    FLASH effect definition & Effect observed in terms of sparing effect and/or tumour control & Text \\
    Euthanasia criteria & Termination criteria (relevant for survival endpoint) & Text \\

    \hline
    \rowcolor{lightgray} \multicolumn{3}{c}{\textbf{Machine-specific parameters**}} \\
    Particle & Proton, Electron, Xray or Heavy Ions & Multi-select \\
    Beam energy [MeV] & Energy of the beam & Number \\
    Beam current [nA] & Current of the beam at target (or, if not given, at the source) & Number \\
    Micro-pulse width* [$\mu$s] & Time between the leading and trailing edges of a single pulse & Number \\
    Micro-pulse frequency* [Hz] & Number of pulses of a repeating signal in a specific time unit & Number \\
    (Macro) pulse width* [$\mu$s] & Time between the leading and trailing edges of a macro pulse & Number \\
    (Macro) pulse frequency* [Hz] & Number of macro pulses of a repeating signal in a specific time unit & Number \\
    Dose per (macro) pulse [Gy/pulse] & Dose divided by the number of delivered macro pulses & Number \\
    Machine type/name & Machine name and Vendor & Multi-select \\
    \multicolumn{3}{c}{\emph{*see Figure \ref{fig: machine param} for details. **If different, one entry for CONV and UHDR separately.}} \\

    \hline
    \rowcolor{lightgray} \multicolumn{3}{c}{\textbf{Dosimetric parameters**}} \\
    Dose [Gy] & Dose (or dose range) delivered to the target & Number \\
    Average dose rate [Gy/s] & Dose rate expressed as total dose divided by the treatment time & Number \\
    PBS or instantaneous dose rate [Gy/s] & PBS dose rate recalculated using the code presented in \ref{DR_code} (for proton PBS delivery) or instantaneous dose rate & Number \\
    Field size [cm2] & Field dimension at target position (as defined by authors) & Number \\
    Dosimetric parameters & Distal and lateral uniformity, SOBP or therapeutic range width & Number \\
    Irradiation type & Passive or active scanning, narrow or broad beam & Multi-select \\
    Experimental setup & Description of the setup, source to target distance, alignment system used, etc. & Text \\
    \multicolumn{3}{c}{\emph{**if different, one entry for CONV and UHDR separately.}} 
    
\label{tab:Parameters}
\end{longtable}

\clearpage
\begin{table}
\scriptsize
\caption{Comparison of the skin score toxicity scale used in all papers where skin toxicity was reported. Light skin toxicities are shown in yellow, moist desquamation-related toxicities are shown in orange, and severe skin toxicities are shown in red.}
    \centering
    \begin{subtable}[t]{0.22\textwidth}
        \vspace{0pt} 
        \centering
        \begin{tabularx}{\linewidth}{>{\raggedright\arraybackslash}X}
            \toprule
            \midrule
            \textbf{\cite{sorensen_pencil_2022}}\\
            \midrule
            \midrule
            \cellcolor{yellow!30} \textbf{0.5} - Normal \\
            \cellcolor{yellow!30} \textbf{1} - Abnormality with reddening \\
            \cellcolor{yellow!30} \textbf{1.5} - Moist desquamation of one small area \\
            \cellcolor{orange!30}\textbf{2.0} - Moist desquamation of 25\% of skin area \\
            \cellcolor{orange!30}\textbf{2.5} - Moist desquamation of 50\% of skin area \\
            \cellcolor{orange!30}\textbf{3.0} - Moist desquamation of 75\% of skin area \\
            \cellcolor{orange!30}\textbf{3.5} - Moist desquamation of the entire area \\
            \midrule
            \bottomrule
        \end{tabularx}
    \end{subtable}%
    \hspace{0.024\textwidth} %
    \begin{subtable}[t]{0.22\textwidth}
        \vspace{0pt} 
        \centering
        \begin{tabularx}{\linewidth}{>{\raggedright\arraybackslash}X}
            \toprule
            \midrule
            \textbf{\cite{mascia_impact_2023}}\\
            \midrule
            \midrule
            \cellcolor{yellow!30}\textbf{1} - Normal \\
            \cellcolor{yellow!30}\textbf{2} - Alopecia \\
            \cellcolor{yellow!30}\textbf{3} - Erythema \\
            \cellcolor{yellow!30}\textbf{4} - Dry desquamation \\
           \cellcolor{orange!30} \textbf{5} - Moist desquamation of 30\% of the skin \\
            \cellcolor{orange!30}\textbf{6} - Moist desquamation of 70\% of the skin \\
            \midrule
            \bottomrule
        \end{tabularx}
    \end{subtable}%
    \hspace{0.024\textwidth}
    \begin{subtable}[t]{0.22\textwidth}
        \vspace{0pt} 
        \centering
        \begin{tabularx}{\linewidth}{>{\raggedright\arraybackslash}X}
            \toprule
            \midrule
            \textbf{\cite{iturri_oxygen_2023}}\\
            \midrule
            \midrule
            \cellcolor{yellow!30}\textbf{0} - Normal \\
            \cellcolor{yellow!30} \textbf{1} - Dull, faint erythema with epilation \\
            \cellcolor{yellow!30}\textbf{2} - Bright erythema with dry desquamation \\
            \cellcolor{orange!30}\textbf{2.5} - Patchy moist desquamation with moderate erythema\\
            \cellcolor{orange!30}\textbf{3} - Confluent moist desquamation with pitting erythema \\
           \cellcolor{red!30}\textbf{4} - Spontaneous bleeding \\
            \cellcolor{red!30}\textbf{4.5} - Ulceration \\
            \cellcolor{red!30}\textbf{5} - Necrosis \\
            \midrule
            \bottomrule
        \end{tabularx}
    \end{subtable}%
    \hspace{0.024\textwidth}
    \begin{subtable}[t]{0.24\textwidth}
        \vspace{0pt} 
        \centering
        \begin{tabularx}{\linewidth}{>{\raggedright\arraybackslash}X}
            \toprule
            \midrule
            \textbf{\cite{velalopoulou_flash_2021}}\\
            \midrule
            \midrule
            \cellcolor{yellow!30}\textbf{0.5} - Normal\\
            \cellcolor{yellow!30}\textbf{0.75} - Slight abnormality\\
            \cellcolor{yellow!30}\textbf{1} - Definite abnormality with reddening\\
            \cellcolor{yellow!30}\textbf{1.25} - Severe reddening\\
            \cellcolor{yellow!30}\textbf{1.5} - Moist breakdown in one very small area\\
            \cellcolor{orange!30}\textbf{1.75} - Moist desquamation in small areas\\
            \cellcolor{orange!30}\textbf{2} - Breakdown of large area, possibly moist in places\\
            \cellcolor{orange!30}\textbf{2.5} - Breakdown of large areas of skin with definite moist exudate\\
            \cellcolor{orange!30}\textbf{3} - Breakdown of most of skin with moist exudate\\
            \cellcolor{red!30}\textbf{3.5} - Complete moist breakdown\\
            \midrule
            \bottomrule
        \end{tabularx}
    \end{subtable}%

    \centering
    \begin{subtable}[t]{0.22\textwidth}
        \vspace{4pt} 
        \centering
        \begin{tabularx}{\linewidth}{>{\raggedright\arraybackslash}X}
            \toprule
            \midrule
            \textbf{\cite{field_effects_1974}}\\
            \midrule
            \midrule
            \cellcolor{yellow!30} \textbf{1} - Definite skin reddening \\
            \cellcolor{yellow!30} \textbf{1.5} - Very slight breakdown of the skin surface \\
            \cellcolor{orange!30} \textbf{2} - Breakdown of a large area and/or toes stuck together, possibly moist in places \\
            \cellcolor{red!30}\textbf{2.5} - About one half or the foot broken down \\
            \cellcolor{red!30}\textbf{3} - Almost complete breakdown of the foot \\
            
            \midrule
            \bottomrule
        \end{tabularx}
    \end{subtable}%
    \hspace{0.024\textwidth} %
    \begin{subtable}[t]{0.22\textwidth}
        \vspace{4pt} 
        \centering
        \begin{tabularx}{\linewidth}{>{\raggedright\arraybackslash}X}
            \toprule
            \midrule
            \textbf{\cite{konradsson_comparable_2022}}\\
            \midrule
            \midrule
            \cellcolor{yellow!30}\textbf{1} - Normal\\
            \cellcolor{yellow!30}\textbf{2} - Hair loss\\
            \cellcolor{yellow!30}\textbf{3} - Erythema\\
            \cellcolor{yellow!30}\textbf{4} - Dry desquamation\\
            \cellcolor{orange!30}\textbf{5} - $<$30\% moist desquamation\\
            \cellcolor{orange!30}\textbf{6} - $>$30\% moist desquamation\\
            \midrule
            \bottomrule
        \end{tabularx}
    \end{subtable}%
    \hspace{0.024\textwidth}
    \begin{subtable}[t]{0.22\textwidth}
        \vspace{4pt} 
        \centering
        \begin{tabularx}{\linewidth}{>{\raggedright\arraybackslash}X}
            \toprule
            \midrule
            \textbf{\cite{soto_flash_2020}}\\
            \midrule
            \midrule
            \cellcolor{yellow!30}\textbf{1} - $<$50\% depigmentation \\
            \cellcolor{yellow!30} \textbf{2} - 50\% depigmentation \\
            \cellcolor{yellow!30}\textbf{3} - 50\% alopecia \\
            \cellcolor{yellow!30}\textbf{4} - $>$50\% alopecia\\
            \cellcolor{orange!30}\textbf{5} - $<$50\% ulceration (alopecia and depigmentation) \\
           \cellcolor{red!30}\textbf{6} -  $>$50\% ulceration (alopecia and depigmentation) \\
            \midrule
            \bottomrule
        \end{tabularx}
    \end{subtable}%
    \hspace{0.024\textwidth}
    \begin{subtable}[t]{0.24\textwidth}
        \vspace{4pt} 
        \centering
        \begin{tabularx}{\linewidth}{>{\raggedright\arraybackslash}X}
            \toprule
            \midrule
            \textbf{\cite{duval_comparison_2023}}\\
            \midrule
            \midrule
            \cellcolor{yellow!30}\textbf{0} - Normal\\
            \cellcolor{yellow!30}\textbf{1} - Dry/Pre moist desquamation\\
            \cellcolor{orange!30}\textbf{2} - Partial thickness epidermal lysis (ulceration)/desquamation with moisture\\
            \cellcolor{red!30}\textbf{3} - Full thickness epidermal lysis (ulceration)/desquamation with moisture\\
            \midrule
            \bottomrule
        \end{tabularx}
    \end{subtable}%

    \centering

     \begin{subtable}[t]{0.22\textwidth}
        \newpage 
        \centering
        \begin{tabularx}{\linewidth}{>{\raggedright\arraybackslash}X}
            \toprule
            \midrule
            \textbf{\cite{rudigkeit_proton-flash_2024}}\\
            \midrule
            \midrule
            \cellcolor{yellow!30}\textbf{0} - Normal\\
            \cellcolor{yellow!30}\textbf{1} - Dry desquamation\\
            \cellcolor{yellow!30}\textbf{2} - Crust formation\\
            \cellcolor{orange!30}\textbf{3} - Moist desquamation\\
    
            \midrule
            \bottomrule
        \end{tabularx}
    \end{subtable}%
     \hspace{0.024\textwidth}
    \begin{subtable}[t]{0.22\textwidth}
    
        \newpage 
        \centering
        \begin{tabularx}{\linewidth}{>{\raggedright\arraybackslash}X}
            \toprule
            \midrule
            \textbf{\cite{tavakkoli_anesthetic_2023}}\\
            \midrule
            \midrule
            \cellcolor{yellow!30}\textbf{0} - Normal \\
            \cellcolor{red!30}\textbf{1} - Ulceration \\
            \midrule
            \bottomrule
        \end{tabularx}
    \end{subtable}%

\label{tab:skin_score_scales}
\end{table}

\clearpage
\section{Figures}

\begin{figure}[h!]
    \centering
    \includegraphics[width=0.7\textwidth]{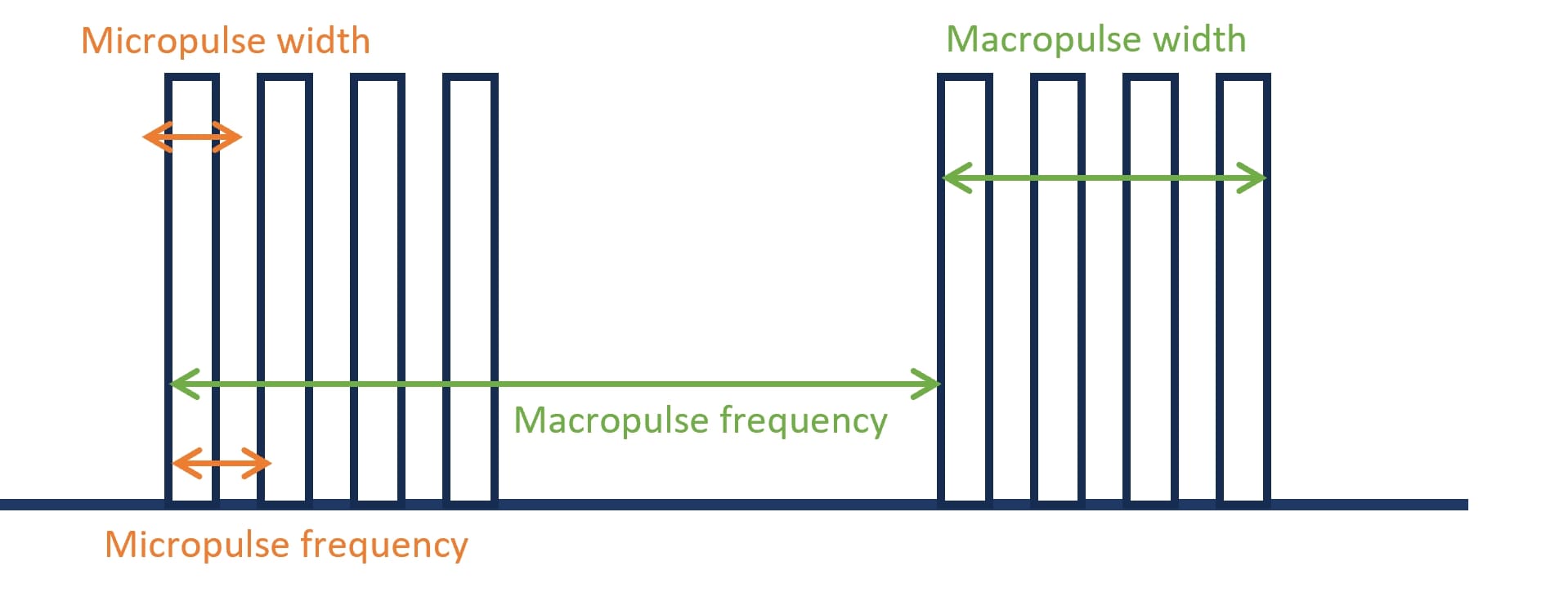}
    \caption{Illustration of the definition used to collect machine specific parameters.}
    \label{fig: machine param}
\end{figure}

\end{document}